# Magnetic Quantum String Waves in Non-Degenerate Quantum Plasma

Levan N.Tsintsadze

Andronikashvili Institute of Physics, Javakhishvili Tbilisi State University, Tbilisi 0128, Georgia

ABSTRACT

Instabilities of transverse waves due to a constant magnetic field in a non-degenerate quantum electron-ion plasmas are studied. A new type of cyclotron oscillation with a small growth rate is disclosed. Excitation of a new quantum mode is also revealed. Furthermore, the kinetic instabilities of Alfven waves are discussed and the growth rates are obtained. A novel branch of waves that we call a magnetic quantum string waves is found.

Recently there has been a great deal of interest in the study of non-degenerate as well as degenerate quantum plasmas. This interest is triggered by its potential application in modern technology, e.g. metallic and semiconductor nanostructures [1-4]. Moreover, the quantum electron-ion or electron-positron-ion magnetoactive plasmas are common in planetary interiors, in compact astrophysical objects [5], as well as in the next generation intense laser-solid density plasma experiments. Various aspects of linear and nonlinear propagation characteristics of electrostatic or electromagnetic modes in the context of isotropic Fermi surfaces have been investigated in numerous papers [6-14]. However, as is well known, the concept of spherical symmetry of the Fermi surface is no more valid in certain systems even in a collisionless regime of Fermi gas [15,16]. Hence, one should bear in mind that the momenta in parallel and perpendicular directions are distinct in considerations of dynamics of non-degenerate or degenerate quantum plasma in the presence of a magnetic field. Precise study in such scenarios demands elongated or even cylindrical Fermi surfaces [15,17].

It is also well known that the magnetic field in a Fermi gas drives some interesting phenomena, such as the change of shape of the Fermi sphere, thermodynamics [18,19], de Haas–van Alphen [20] effect, skin effect, etc. Recently an adiabatic magnetization process has been proposed in Ref.[21] for cooling the Fermi quantum plasma, whereas quantum Weibel instabilities have been reported in Refs.[22,23]. Employing the quantum kinetic equation derived in Ref.[10] the dispersion properties of various modes have been discussed in quantum plasmas [10-14,24].

In this Letter, we consider instabilities of transverse waves due to a constant magnetic field in the non-degenerate quantum electron-ion plasmas.

As is well known, in classical physics the energy of charge does not change in a homogeneous and time independent magnetic field. However, as was shown by Landau [15], the situation drastically changes in quantum mechanics. Namely, in the constant magnetic field, $\vec{H}(0,0,H_0)$, the quantization of the orbital motion of particles leads to change of energy of particles. Perpendicular component of momentum relative to a direction of the magnetic field is defined by the strength of the magnetic field, but not a temperature. Taking into account the quantization of energy of particles, Landau derived a quasi-classical distribution function for the non-degenerate particles (also see Kelly [25])

$$F_\alpha = n_\alpha \cdot f_\alpha = n_\alpha \cdot \left(\frac{m_\alpha}{2\pi}\right)^{\frac{3}{2}} \frac{1}{T_\parallel^{1/2} \cdot \varepsilon_{\perp\alpha}} exp\left(-\frac{m_\alpha v_\perp^2}{2\varepsilon_{\perp\alpha}} - \frac{m_\alpha v_\parallel^2}{2T_\parallel}\right), \quad (1)$$

where $\varepsilon_{\perp\alpha} = \frac{\hbar\omega_{c\alpha}}{2} \cdot coth\frac{\hbar\omega_{c\alpha}}{2T_\parallel}$, $v_\perp^2 = v_x^2 + v_y^2$, $\omega_{c\alpha} = \frac{e_\alpha H_0}{m_\alpha c}$ - is the cyclotron frequencies of particles, $H_0$ is the constant magnetic field directed along the z-axis. Note that for $\hbar\omega_{c\alpha} \gg T_\parallel$, $\varepsilon_{\perp\alpha}$ equals $\frac{\hbar\omega_{c\alpha}}{2}$, thus in such case the distribution function (1) is strongly anisotropic.

We first study the linear Weibel instabilities. To this end, we employ the linear dispersion relation of a right-hand polarized transverse electromagnetic waves propagating along the z-axis [26]

$$\frac{k^2 c^2}{\omega^2} = 1 - \frac{\omega_{pe}^2}{\omega^2} \int \frac{d\vec{v} \cdot \frac{v_\perp}{2}}{\omega - \omega_{ce} - kv_\parallel} \left[(\omega - kv_\parallel)\frac{\partial}{\partial v_\perp} + kv_\perp \frac{\partial}{\partial v_\parallel}\right] f_e, \quad (2)$$

where $\omega_{pe} = \left(\frac{4\pi e^2 n_{0e}}{m_e}\right)^{1/2}$ and ions are assumed to be immobile.

Substituting the distribution function (1) for electrons into Eq.(2), after integration we obtain

$$\frac{k^2 c^2}{\omega^2} = 1 + \frac{\omega_{pe}^2}{\omega^2}\left\{\eta - \left(\eta + \frac{\omega}{\omega - \omega_{ce}}\right)I_+\left(\frac{\omega - \omega_{ce}}{ku_\parallel}\right)\right\}, \quad (3)$$

where $I_+(x) = xe^{-x^2/2} \int_{+i\infty}^{x} d\tau e^{-\tau^2/2}$ [27],

$\eta = \frac{\varepsilon_\perp}{T_\parallel} - 1$, and $u_\parallel = \sqrt{\frac{T_\parallel}{m_e}}$ is the thermal velocity along the magnetic field.

We recall an asymptotic expansion of the function $I_+(x)$ [27], which for the case $|x| \gg 1$, $|Re(x)| \gg |Im(x)|$ reads

$$I_+(x) = 1 + \frac{1}{x^2} + \frac{3}{x^4} + \cdots - i\sqrt{\frac{\pi}{2}} \cdot x \cdot e^{-x^2/2} \tag{4}$$

and for $|x| \ll 1$, $|Re(x)| \gg |Im(x)|$ is

$$I_+(x) = -i\sqrt{\frac{\pi}{2}} \cdot x. \tag{5}$$

We now consider the cyclotron waves near an electron cyclotron frequency, i.e. $|\omega - \omega_{ce}| \ll ku_\parallel$. As is well known, in this range of frequency at $T_\parallel = \varepsilon_\perp = T$ waves are strongly damped by the Landau mechanism. Here we will show that if $\varepsilon_\perp \gg T_\parallel$, then even for $|\omega - \omega_{ce}| \ll ku_\parallel$, exists the possibility of cyclotron wave propagation with the real frequency larger than the imaginary one. To this end, we use Eq.(5) of the function $I_+\left(\frac{\omega-\omega_{ce}}{ku_\parallel}\right)$ and assume that $k^2 c^2 \gg \omega^2$ to obtain

$$\omega = \omega_{ce}\left(1 - \frac{T_\parallel}{\varepsilon_\perp}\right) - i\sqrt{\frac{2}{\pi}}\frac{T_\parallel}{\varepsilon_\perp}\left(\frac{k^2 c^2}{\omega_{pe}^2} - \eta\right) ku_\parallel. \tag{6}$$

From which follow expressions of the real and imaginary frequencies as

$$Re\omega = \omega_{ce}\left(1 - \frac{T_\parallel}{\varepsilon_\perp}\right), \tag{7}$$

$$Im\omega = -\sqrt{\frac{2}{\pi}} \cdot \frac{T_\parallel}{\varepsilon_\perp} \cdot \left(\frac{k^2 c^2}{\omega_{pe}^2} - \eta\right) ku_\parallel. \tag{8}$$

It should be noted that the both expressions (7) and (8) are novel and demonstrate that the anisotropic distribution function not only leads to instabilities, the result of which is the growth of the amplitude of electromagnetic waves in time, but also a new cyclotron oscillations arise as reveals Eq.(7). The excitation of cyclotron oscillations takes place at $\frac{\varepsilon_\perp}{T_\parallel} > 1 + \frac{k^2 c^2}{\omega_{pe}^2}$ as shows Eq.(8), i.e. this is a new branch of oscillation due to the quantization of energy of electrons.

Next, we consider the cyclotron waves near the electron cyclotron frequency with the condition $|\omega - \omega_{ce}| \gg ku_\parallel$, which means that the frequency lies outside the resonance absorption line. In this case use of the asymptotic expansion (4) for the function $I_+(x)$ in Eq.(3) yields

$$\frac{k^2 c^2}{\omega^2} = 1 - \frac{\omega_{pe}^2}{\omega^2}\left[\left(\frac{\omega}{\omega - \omega_{ce}}\right) + \frac{k^2 u_\parallel^2}{(\omega - \omega_{ce})^2} + \frac{k^2 u_\parallel^2 \omega_{ce}}{(\omega - \omega_{ce})^3}\right] + i\sqrt{\frac{\pi}{2}}\frac{\omega_{pe}^2}{\omega ku_\parallel}\left[(\eta + 1) - \eta\frac{\omega_{ce}}{\omega}\right]e^{-\frac{(\omega-\omega_{ce})^2}{k^2 u_\parallel^2}}, \tag{9}$$

where $u_\perp^2 = \frac{2\varepsilon_\perp}{m_e}$.

Note that Eq.(9) is enriched with two new terms owing to the quantization of energy of electrons.

In the following we investigate the equation (9) for different ranges of frequencies. Let us first neglect the last imaginary term in Eq. (9). Then under condition $\omega \gg \omega_{ce}$, we obtain

$$\frac{k^2 c^2}{\omega^2} = 1 - \frac{\omega_{pe}^2}{\omega^2}\left(1 + \frac{k^2 u_\perp^2}{\omega^2}\right). \tag{10}$$

This dispersion equation was first derived by Weibel, solutions of which are

$$\omega_1^2 = \omega_{pe}^2 + k^2 c^2, \quad \omega_2^2 = -\frac{\omega_{pe}^2 k^2 u_\perp^2}{\omega_{pe}^2 + k^2 c^2}. \tag{11}$$

The frequency $\omega_2$ describes two types of hydrodynamic instabilities

$$\omega_2 = i\frac{u_\perp}{c}\omega_{pe}, \quad \omega_{pe}^2 \ll k^2 c^2 \tag{12}$$

$$\omega_2 = ik\sqrt{\frac{\hbar \omega_{ce}}{m_e}}, \quad \omega_{pe}^2 \gg k^2 c^2 \tag{13}$$

Next, in the case of the helical waves, $\omega \ll \omega_{ce}$, we consider the kinetic instability. Assuming $\omega = \omega' + i\omega''$, from Eq.(9) for the $Re\omega = \omega'$ we obtain the modified helical dispersion equation

$$\omega' = \frac{k^2 c^2}{\omega_{pe}^2}\left(1 + \frac{u_\perp^2}{c^2}\frac{\omega_{pe}^2}{\omega_{ce}^2}\right)\omega_{ce}, \tag{14}$$

and for the growth rate $\omega''$ we have

$$\omega'' = \sqrt{\frac{\pi}{2}}\frac{\omega' \omega_{ce}}{k u_\parallel}\left(\frac{\omega_{ce}}{\omega'}\eta - \frac{\varepsilon_\perp}{T_\parallel}\right)e^{-\frac{\omega_{ce}^2}{k^2 u_\parallel^2}}. \tag{15}$$

We note here that the equation (14) has two interesting solutions. One is usual helical waves $\omega' = \frac{k^2 c^2}{\omega_{pe}^2}\omega_{ce}$, for which the inequality $u_\perp^2 \omega_{pe}^2 \ll c^2 \omega_{ce}^2$ should be satisfied. The second branch of waves is due to quantization of energy of electrons and at $u_\perp^2 \omega_{pe}^2 \gg c^2 \omega_{ce}^2$, we derive a new quantum mode of the transverse electromagnetic waves

$$\omega' = \frac{k^2 u_\perp^2}{\omega_{ce}} = \frac{\hbar k^2}{m_e}, \tag{16}$$

which is twice larger than the frequency of quantum oscillations of a free electron of the longitudinal waves, $\omega_l = \frac{\hbar k^2}{2 m_e}$ [10].

The growth rate of this new mode (16) is

$$\omega'' = \sqrt{\frac{\pi}{2}}\frac{\omega_{ce}^2}{k u_\parallel}\eta\, e^{-\frac{\omega_{ce}^2}{2k^2 u_\parallel^2}}. \tag{17}$$

We now consider the kinetic instabilities of the Alfven waves assuming that ions are cold. Dispersion equation for the right-hand polarized wave reads

$$\frac{k^2 c^2}{\omega^2} = \varepsilon_x^e - i\varepsilon_y^e + \varepsilon_x^i + i\varepsilon_y^i, \quad \text{or}$$

$$\frac{k^2 c^2}{\omega^2} = 1 - \frac{\omega_{pe}^2}{\omega^2}\left[-\eta + \left(\eta + \frac{\omega}{\omega - \omega_{ce}}\right) I_+\left(\frac{\omega - \omega_{ce}}{k u_{\|e}}\right)\right] - \frac{\omega_{pi}^2}{\omega(\omega - \omega_{ci})}, \quad (18)$$

where $\omega_{pi}^2 = \frac{4\pi e^2 n_0}{m_i}$ and $\omega_{ci} = \frac{eH_0}{m_i c}$ are the Langmuir and cyclotron frequencies of ions, respectively.

In the range of frequencies $\omega \ll \omega_{ci}$ and $|\omega \pm \omega_{ci}| \gg k u_\|$ we obtain the following expression for a real part of frequency

$$Re\omega^2 = k^2 V_{A_i}^2 \left[1 + \frac{u_\perp^2}{V_{A_e}^2}\left(1 - \frac{T_\|}{\varepsilon_\perp}\right)\right], \quad (19)$$

and for the growth rate we have

$$Im\omega = \sqrt{\frac{\pi}{8}} \frac{\omega_{pe}^2}{Re\omega \cdot k u_\|} \cdot \frac{V_{A_i}^2}{c^2} \omega_{ce} \left(\eta - \frac{\omega}{\omega_{ce}}\right) e^{-\frac{\omega_{ce}^2}{2k^2 u_\|^2}}, \quad (20)$$

where $V_{A_\alpha} = \frac{H_0}{\sqrt{4\pi m_\alpha n_0}}$ is the Alfven velocity.

Note that if $T_\| \gg \hbar \omega_{ce}$, the expression (20) becomes the damping decrement ($\eta = 0$) of Alfven waves.

It should be emphasized that if $V_{A_e}^2 \gg u_\perp^2 = \frac{2\varepsilon_\perp}{m_e} \simeq \frac{\hbar \omega_{ce}}{m_e}$, then equations (19) and (20) describe Alfven waves, whereas in the opposite case $u_\perp^2 \gg V_{A_e}^2$ and $\hbar \omega_{ce} \gg T_\|$, we uncover a new type of waves

$$Re\omega^2 = \frac{\hbar \omega_{ce}}{m_i}\left(1 - \frac{T_\|}{\varepsilon_\perp}\right) \cdot k^2 \quad (21)$$

and a new type of velocity as well

$$c_0 = \sqrt{\frac{\hbar \omega_{ce}}{m_i}\left(1 - \frac{T_\|}{\varepsilon_\perp}\right)} \simeq \sqrt{\frac{\hbar \omega_{ce}}{m_i}}.$$

We specifically note here that these novel waves, which we call a magnetic quantum string waves arise because of anisotropy of the distribution function [28].

In summary, we have investigated instabilities of non-degenerate quantum plasma due to a constant magnetic field. Studying the electron oscillations we have shown that exist a new cyclotron oscillations $\omega \simeq \omega_{ce}$ with a small growth rate. In the frequency range $\omega_{ce} \gg \omega$ in addition to the helical waves, we have disclosed the excitation of quantum oscillations $\omega = \frac{\hbar k^2}{m_e}$ owing to the quantization of energy of electrons. Moreover, we have considered the kinetic instabilities of the Alfven waves and defined the growth rate due to the magnetic field. Finally, we have uncovered a new branch of waves $\omega = k\sqrt{\frac{\hbar \omega_{ce}}{m_i}}$, which we have named as the magnetic quantum string waves. These investigations may play an essential role for the description of complex phenomena that appear in astrophysical objects, as well as in high intense laser-matter interactions.